# The Interplay of Topological Surface and Bulk Electronic States in $Bi_2Se_3$

Megan Romanowich[1], Mal-Soon Lee[1], Duck-Young Chung[2], Jung-Hwan Song[2], S.D. Mahanti[1], Mercouri G. Kanatzidis[2,3], Stuart H. Tessmer[1]

[1]Department of Physics & Astronomy, Michigan State University, East Lansing, Michigan 48824
[2]Materials Science Division, Argonne National Laboratory, Argonne, Illinois 60439
[3]Department of Chemistry, Northwestern University, Evanston, Illinois 60208

Abstract

In this Letter we present scanning tunneling microscopy density-of-states measurements and electronic structure calculations of the topological insulator $Bi_2Se_3$. The measurements show significant background states in addition to the expected Dirac cone. Density functional calculations using a slab model and analysis of the partial density-of-states show that the background is consistent with bulk-like states with small amplitudes at the surface. The topological surface states coexist with bulk-like states in the valence band, appearing as a shoulder in the projected band structure. These results strongly support the picture suggested by recent scattering experiments of the quantum interference of topological and bulk-like surface states.

The discovery of topological insulators (TI) has transformed our understanding of insulating behavior[1]. These novel systems are insulators in the bulk while conducting along the surface, exhibiting topologically-protected gapless surface states (TSS) with striking spin textures [2, 3]. This remarkable electronic structure arises from spin-orbit interactions and time reversal invariance, resulting in the formation of massless Dirac cones in the single particle energy dispersion [1].

Bismuth-selenide, $Bi_2Se_3$, is a layered 3-dimensional material that exhibits topological simplicity, typified by a single Dirac cone forming the TSS [4-6]; moreover, it is readily cleavable, allowing a contamination-free and chemically-inert surface to be exposed for experimental probes. For these reasons, $Bi_2Se_3$ represents a model topological insulator. Because the surface state primarily resides in the top layer, many research groups employ surface probe methods of angle-resolved photo-emission spectroscopy (ARPES) [6-12] and scanning tunneling microscopy (STM) [2,3,5,13-17] to examine the nature of the TSS. Moreover, surprisingly, a recent scattering experiment has shown that bulk electronic states coexist in the same energy range as the TSS and can therefore interfere with this state [16].

Multiple spectroscopic studies of $Bi_2Se_3$ and other TI materials have probed the TSS in proximity to surface defects, making significant progress in our understanding of these interactions [2,3]. In addition, many groups [2,15,17-19] observe a spectroscopic feature not clearly understood; the local density of states (LDOS) at the surface does not vanish at the Dirac point (DP) as expected from the linear dispersion of the TSS near the DP. Instead, a non-zero "background" LDOS is consistently observed. The previous studies do not offer a satisfactory explanation of this feature leaving open the possibility that it is an artifact of non-ideal samples. In this Letter, we focus on the physics of this background feature through STM-based tunneling spectroscopy measurements and electronic structure calculations. The experimental approach employs careful sample preparation and high quality spectroscopy so that we are confident that we are probing electronic structure intrinsic to $Bi_2Se_3$. The theoretical studies involve a detailed study of the surface and bulk electronic structures using a slab model where the slab thickness is varied systematically.

Bismuth-selenide is ideal for surface studies. Five atomic planes with atomic order Se1-Bi1-Se2-Bi1-Se1 form a quintuple layer (QL); the QLs are weakly bound to each other, making it possible to readily expose a pristine surface for study. Fig. 1(a) shows a 6-QL slab. Differential conductance maps employing scanning tunneling microscopy and spectroscopy methods reveal the scattering of the TSS from defects [16,17]; enhancement in the quasiparticle interference patterns demonstrate mixing of the scattered TSS with the bulk continuum in precisely the direction in **k**-space where band structure calculations [4,6,16,20] show a bulk valence band peak overlapping the energy of the DP. Therefore, although $Bi_2Se_3$ has a simple Dirac cone representing its TSS, bulk continuum states appear to contribute to differential conductance measurements. The motivation of our study is to explore and understand the interplay of the bulk states with the surface using both experiment (STM) and theory (*ab initio* electronic structure calculations).

Samples of $Bi_2Se_3$ were prepared by slowly cooling an elemental mixture from 850°C and X-ray diffraction analysis showed no significant impurities. Stoichiometric $Bi_2Se_3$ is typically n-type, resulting in a shift of the Fermi level so that it is near the gap edge; this is believed to result from Se vacancies in the crystal [21]. For this experiment, in addition to stoichiometric samples, we probed samples with excess Bi which tends to restore the Fermi level towards the center of the band gap. We cleaved the samples in a nitrogen environment in a custom-built integrated processing chamber and transferred them onto a cryogenic scanning tunneling microscope without exposure to air. Prior to cooling to liquid He temperature, the measurement space is evacuated to $3 \times 10^{-6}$ torr, before backfilling with He exchange gas to minimize surface contaminants. We have found that these procedures result in exceptionally clean samples and stable STM spectroscopy.

Atomic resolution images of the surface from STM at 4.2K show the surface Se layer of the crystal as shown in the lower inset of Fig. 1(b). While some areas showed characteristic Bi substitutional defects [3], there were many areas free of defects, allowing scanning tunneling spectroscopy of the pristine $Bi_2Se_3$ surface state. Fig. 1(b) shows a representative current-voltage curve acquired with the STM tip positioned above an area free of defects. In all tunneling measurements presented here, the bias voltage is applied to the sample with the tip potential maintained at ground. The top inset presents the expanded I-V in the vicinity of the Dirac cone, clearly showing a nonzero slope around zero applied voltage. At low temperature, the DOS is essentially the slope of the I-V curve; hence we can conclude that there are states in the bulk gap even at the Fermi level (zero bias voltage).

Fig. 1 (c) shows a representative differential-conductance curve obtained on a sample doped with 4% excess Bi. Curves obtained on stoichiometric samples showed similar structure. As a signature of a topological insulator, the spectrum is linear in the range of sample bias voltage +/-0.15 V, forming a "V" shape with the Dirac point at the vertex. We see that for this sample, at locations several nanometers away from defects, the Fermi level (i.e. zero bias voltage) is at the Dirac point. We note that other non-linear features sometimes appear above +0.2 V; their physical origin will be discussed in a forthcoming paper.

The data for Fig. 1 (b) and 1(c) curves were obtained at a temperature of 4.2 K by superposing a sinusoidal voltage of amplitude 4 mV rms with the bias voltage; a lock-in amplifier was used to acquire the dI/dV curve as the voltage was slowly ramped [22]. The displayed curve is the average of 150 curves, acquired successively at the same location to reduce noise in the measurement. The focus of this measurement is the background feature, seen as the vertical offset of the Dirac point. For STM measurements, the magnitude of spectroscopy curves scales with the set point current. Hence, to quantitatively discuss the measured magnitude of the background offset, we need to compare it to other features in the curve. Our approach is to compare it to the Dirac cone plus background states observed at -0.10 V, labeled $\beta$ in the inset of Fig. 1(c). (Of course, any voltage level within the Dirac cone would suffice. The choice of -0.10 V is sufficiently far from the Dirac point and clearly on the linear part of the curve.) Correspondingly, the offset height is labeled $\alpha$ in Fig 1(c) inset. We find that the ratio $\beta/\alpha = 2.87$. Identical measurements at three different locations on this sample yield ratios of $\beta/\alpha$ that agree to within 15%; this level of agreement is roughly consistent with the uncertainty in the measurements, as estimated by the scatter in the data. The reproducibility of this local measurement at different defect-free locations suggests that the offset is not caused by defects or impurities; we believe this experimental evidence strongly suggests that this background is intrinsic to the system.

In order to understand the origin of the non-zero DOS at the DP seen in the tunneling measurements, we look at the electronic structure of a series of 6-18 QLs (30-90 atomic planes). Scalar relativistic electronic structure calculations were carried out within density-functional theory (DFT) using the projector augmented wave (PAW) methods [23] as implemented in the VASP package [24]. The Perdew-Burke-Ernzerhof (PBE) generalized gradient corrected exchange-correlation functional [25] were used in our calculation. An energy cutoff of 400 eV was used for the plane-wave expansion, with a total energy convergence of the order of $10^{-4}$ eV. Spin-orbit interactions (which are essential to get the Dirac cone states) have been included using second variational approximation [26]. Another motivation for our theoretical calculations is to look at the nature of the wave functions of the Dirac cone states and the states associated with the broad valence band maxima (VBM) by calculating the spatially projected DOS. This can tell us whether the states associated with VBM have sufficient amplitudes to be seen in tunneling measurements.

There is already some evidence of nonzero DOS at the DP in earlier electronic structure calculations using slab models and similar approximation (GGA) by Xia *et al.* (12 QLs) [6], Yazyev *et al.* (1-5 QLs) [20], and Kim *et al.* (slab of 7 QLs) [16]; all show that there is a broad peak along the $\overline{\Gamma M}$ direction in the projected bulk band structure. The VBM of this projected band structure lies slightly above the DP. Yazyev *et al.* [20] suggest that the VBM is 40 meV above the DP whereas the other two papers do not give a definite number. There is a basic problem in getting this number as one has to align the slab bands with those obtained from the projected bulk bands. Yazyev *et al.* make this alignment by matching the potential in the "bulk-like" region of their 5-QL slab to the bulk potential. The other two papers do not specify how they have aligned the slab calculations with the bulk projections. In this paper we take two different approaches. One, we look at the energies of different bands at the Γ point below the DP, analyze their orbital characters and match the energies of the slab bands with those of the projected bulk bands of similar orbital structure (in contrast to matching the potential as was done in ref. 20). Second, we avoid this matching problem by studying how the band structure in the slab geometry evolves as we increase the number of QLs from 6 to 18, carefully monitoring the energy difference between the DP and the broad VBM along the $\overline{\Gamma}\overline{M}$ direction. Using these energy differences we extrapolate what will happen for a realistic surface (number of QLs approaching infinity) [27].

Fig. 2(a) shows the band structure along $\overline{K} - \overline{\Gamma} - \overline{M}$ direction for the 6-QL slab (blue lines). In the figure, we also plot the projected bulk band structure with the energy matching criterion discussed in the last paragraph. It can be seen that for the 6QL slab, there is a shoulder near $\overline{\Gamma}$ towards $\overline{M}$, whose maximum energy (peak) is ~0.03 eV lower than the DP. However this peak energy for the projected bulk band structure lies above the DP, in agreement with previous studies [6,16,20] with small differences in the energy difference between the DP and the bulk VBM and the shape of the shoulder. The small energy difference depends on how one aligns the band structures of the slab and bulk as we have discussed earlier. The shape difference is mainly due to different ionic positions. We obtained the band structure with relaxed ionic positions while the earlier studies used experimental structure [28]. In Fig. 2(b) we show how the top most valence band changes as we increase the number of QLs from 6 to 18. Interestingly, when we increase the number of QLs, the shoulder states shift to higher energies, hinting at their bulk origin [29]. With 18QLs, the energy of the top of the shoulder states is slightly above the DP. Fig. 2(c) shows the energy difference between the DP and the top of the shoulder states as a function of the number of QLs. The figure strongly suggests that the change in energy would become even more negative and eventually approach the value corresponding to a bulk with a single surface. Unfortunately due to computer time limitations we could not go beyond 18 QLs. One more point in Fig. 2(c) (for 21 QLs) would have helped us in carrying out a reliable extrapolation. However there is a strong indication from our calculations and earlier estimations that the DP lies below the VBM. We suggest that these shoulder states near the VBM cause the background in the local density of states seen in tunneling measurements.

Because tunneling decays exponentially, scanning tunneling microscopy or spectroscopy measurements are most sensitive to electronic states localized on the surface of a sample. In particular, the nature of tunneling dictates that the topmost selenium atoms (Se1) participate the most in tunneling measurements although Bi1 atoms just below it may also contribute non-negligibly. In order to have a deeper understanding of the spatial characteristics of states associated with different parts of the spectrum (those associated with the Dirac cone and VBM respectively) we choose two regions of the k space where each of these states are predominantly located: Region I corresponds to states near the Γ point (k < 0.015 Å$^{-1}$) focusing on states near the Dirac cone and Region II (0.015 < k < 0.025 Å$^{-1}$) focusing on the states near the VBM (See inset of Fig. 3(a)). Again due to computing time limitations, we look at the 6 QL slab.

Fig. 3 shows the partial density of states (PDOS) for the 6 QL slab associated with Se1 and Bi1 of the top QL (surface) and the middle QL (bulk-like) for the two different regions. The dominant contribution of the states to Region I (corresponding to near) with energies near the DP comes from the Se1 and Bi1 of the top QL, which clearly indicates the surface localization of the Dirac cone states. However, in this region, PDOS with E < -0.05 eV, i.e states away from the DP comes from the middle QLs with dominantly Bi1 character, indicating that these are bulk-like states. The PDOS from Region II comes solely from the shoulder states where the peak is at about -0.03 eV (See the inset of Fig. 3(a)), 0.03 eV below the DP. In agreement with the bulk calculations from Xia *et al.* [4] and Kim *et al.* [16], the shoulder states in our slab calculations are predominantly located in the middle QL (Fig. 3(c) and (d)). The large peak in the Se1 PDOS associated with Region II below 0.0 eV (DP) in the middle QLs indicate that they are clearly bulk-like states. As can be seen, these shoulder states are dominated by Se1 atoms in the middle QLs. Though the shoulder states are primarily concentrated in the bulk region, they have finite amplitudes at the surface Se1 and Bi1 layers and will contribute to the tunneling current. If we increase the number of QLs these shoulder states will overlap in energy with the states near the DP and will contribute to the tunneling current.

In summary, we present high-quality experimental STM density-of-states measurements of the model topological insulating material, Bi$_2$Se$_3$. The data consistently show a background of states in addition to the Dirac cone states. Through electronic structure calculations using slab model and analysis of the partial density-of-states in different quintuple layers, we have shown that the background in the density-of-states is due to bulk-like states with small amplitudes at the surface. The topological surface states coexist in energy range with bulk-like states in the valence band, appearing as "shoulder" states in the projected band structure along the $\overline{\Gamma}\overline{M}$ direction-VBM states. We conclude that due to the large DOS associated with these VBM states one clearly sees them in STM spectroscopic measurements even if the amplitudes of surface atoms associated with these states are quite small, giving rise to the finite offset seen in the dI/dV measurements near the DP, as presented in Fig. 1. Although our data were acquired far from defects and do not involve scattering, the picture we present strongly supports the explanation of the recent scattering experiment of Kim *et al.* [16] in terms of the quantum interference of the TSS and the bulk-like surface state.

**Acknowledgements**

This work was supported by the National Science Foundation, NSF DMR, Grant No. 0906939; the work at Argonne National Laboratory was supported by UChicago Argonne, a U.S. DOE Office of Science Laboratory, operated under Contract No. DE-AC02-06CH11357. M-SL acknowledges support from the Michigan State University College of Natural Science and Department of Physics and Astronomy.

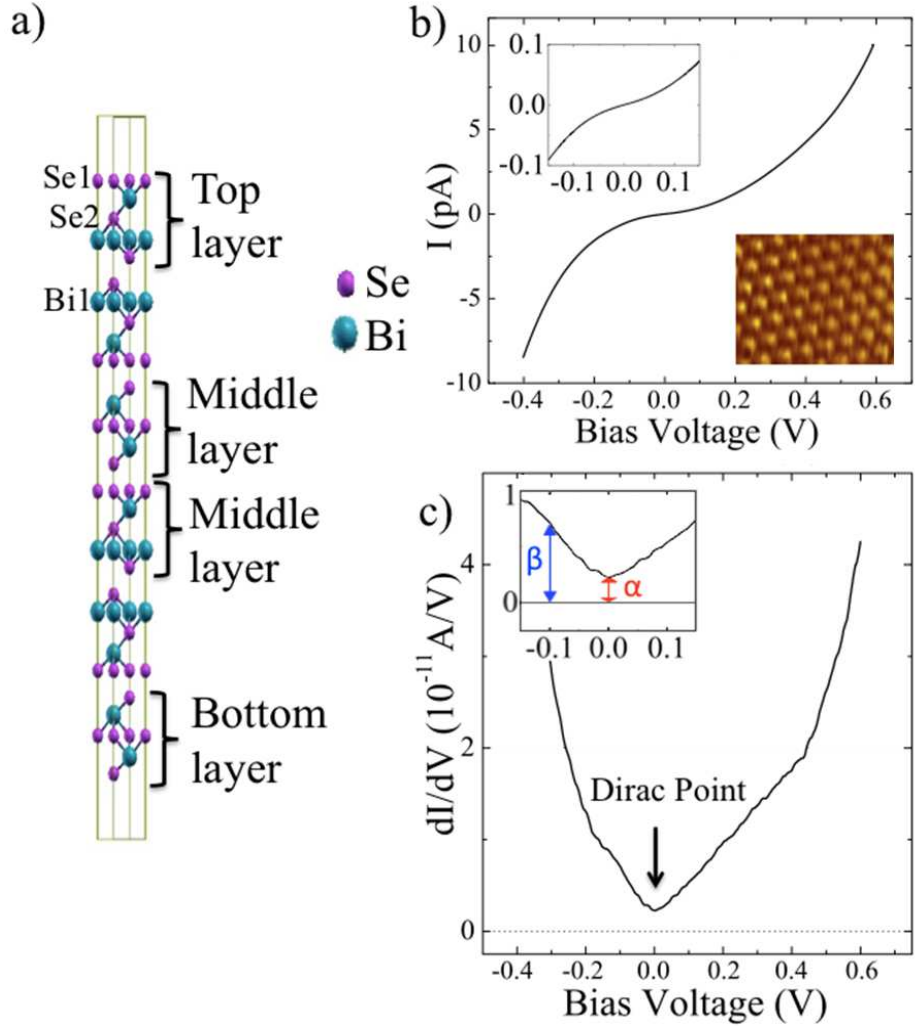

FIG 1. (a) A slab of 6 quintuplet layers of $Bi_2Se_3$ [Ref. 28]. (b) STM tunneling current versus voltage curve of $Bi_{2.04}Se_{2.96}$ acquired at a temperature of 4.2 K. Lower inset: STM topographical image of $Bi_{2.08}Se_{2.92}$ obtained at a bias voltage of -0.6 V and 1.0 nA constant current. Upper inset: Expanded view of the I-V curve in the linear region around the Dirac point; the vertical and horizontal axes are the same as the larger plot. We see a non-zero slope indicating electronic states in the gap, even at the Fermi level, i.e., zero bias voltage. (c) dI/dV of $Bi_{2.04}Se_{2.96}$ at 4.2 K, with the Dirac Point indicated. Inset: An expanded view of the DOS around the Dirac point, showing the offset from zero; the vertical and horizontal axes are the same as the larger plot. The red line, labeled α, denotes the magnitude of the offset, whereas the blue lines, labeled β, denotes a reference height within the Dirac cone. The ratio of β/α acquired at three separate sample locations agree to within experimental uncertainty. Based on the scatter in the data, the noise level is 0.3 pA/V.

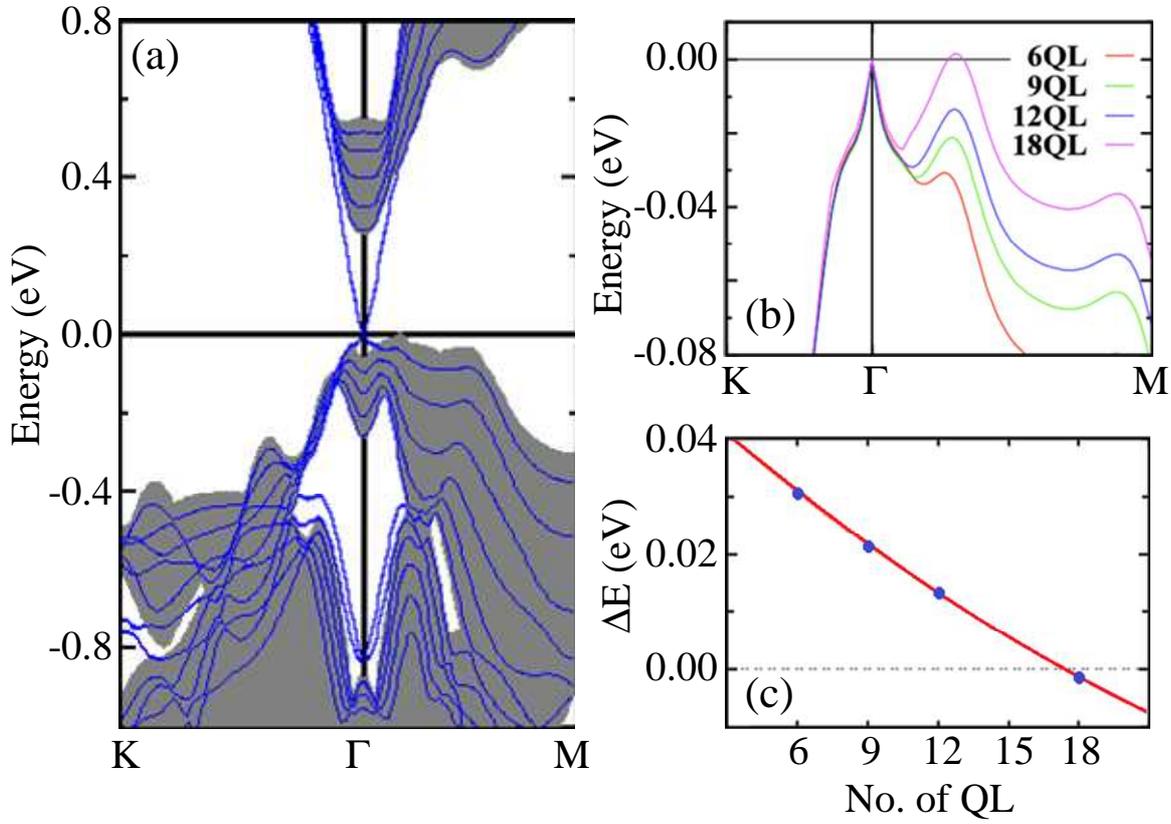

FIG. 2. (a) Band structure of 6 QLs (blue solid lines) along with projected bulk band structure (shaded area). The slab and projected bulk bands were aligned by matching energies of slab and projected bulk bands at the Γ point with similar orbital characters (at energies -0.21 and -0.88 eV below the DP). (b) Band structure near the Dirac cone with different QLs, where the zero of energy is chosen to be at the Dirac Point (DP). (c) Energy difference (ΔE) between the DP and the peak of the shoulder along the $\bar{\Gamma} - \bar{M}$ direction (VBM).

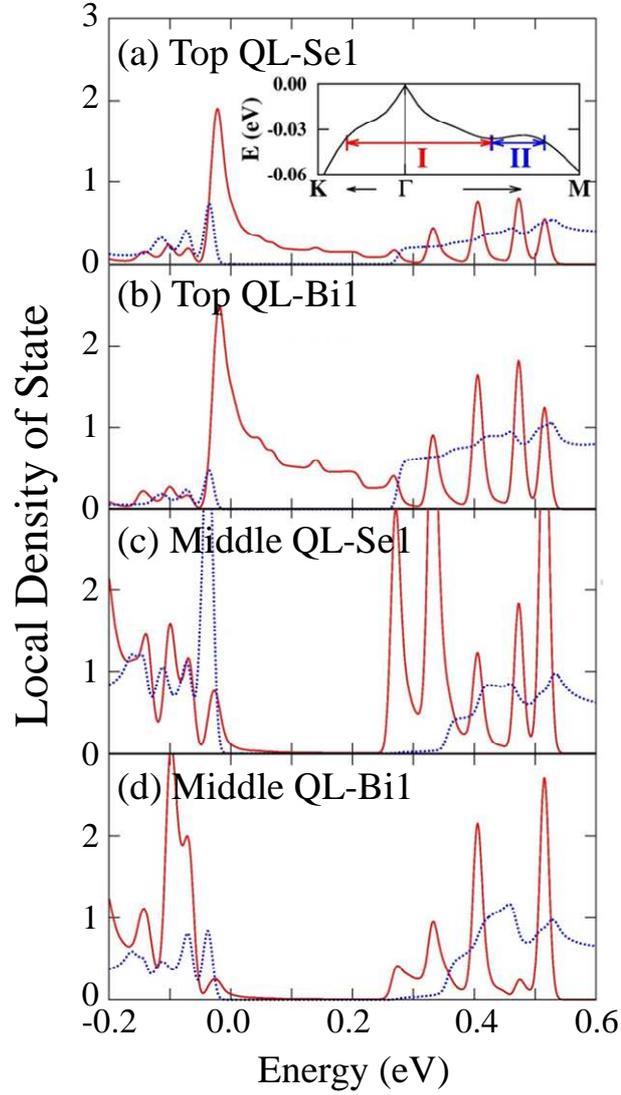

FIG 3: PDOS of Se1 and Bi1 at different QL layers contributed from the two regions in *k*-space; Region I near the $\Gamma$ point (red solid line) where the Dirac point is located, and Region II (blue dotted line) near the peak of the shoulder along $\bar{\Gamma} - \bar{M}$ (VBM). The inset shows the choice of Region I and II, corresponding to the area related to the Dirac cone and that of the shoulder states, respectively. Zero energy in all of these plots is at the Dirac point.